\documentclass[letterpaper, 10 pt, conference, one column, draftcls]{ieeeconf}  
\usepackage{graphicx}
\usepackage{amsmath}
\usepackage{verbatim}
\usepackage{xcolor}
\usepackage[normalem]{ulem}
\usepackage{soul}

\newcommand{\TY}{\color{blue}}

\IEEEoverridecommandlockouts                              

\title{\LARGE \bf
A Service-oriented Metro Traffic Regulation Method for Improving Operation Performance
}

\author{Jiate Luo$^1$, Yin Tong$^{1,*}$, Graziana Cavone$^{2}$, and Mariagrazia Dotoli$^{2}$
\thanks{This work was supported by the National Natural Science Foundation of China under Grant No. 61950410604 and No. 61803317.}
\thanks{$^{1}$J. Luo and Y. Tong (*Corresponding Author) are with the School of Information Science and Technology, Southwest Jiaotong University, Chengdu 611756, China {\tt\small jiateLuo@my.swjtu.edu.cn; yintong@swjtu.edu.cn}}%
\thanks{$^{2}$ G. Cavone and M. Dotoli are with the Department of Electrical and Information Engineering, Polytechnic of Bari, Bari 70125, Italy {\tt\small graziana.cavone@poliba.it; mariagrazia.dotoli@poliba.it}}%
}

\begin{document}

\maketitle
\thispagestyle{empty}
\pagestyle{empty}

\begin{abstract}
For high density metro traffic, nowadays the time-variant passenger flow is the main cause of train delays and stranded passengers. Typically, the main objective of automatic metro traffic regulation methods is to minimize the delay time of trains while passengers' satisfaction is not considered. Instead, in this work a novel framework that integrates a passenger flow module (PFM) and a train operation module (TOM) is proposed with the aim of simultaneously minimizing traffic delays and passengers' discomfort. In particular, the PFM is devoted to the optimization of the headway time in case of platforms overcrowding, so as to reduce the passengers waiting time at platforms and increase the load rate of trains; while the TOM is devoted to the minimization of trains' delays. The two modules interact with each other so that the headway time is automatically adjusted when a platform is overcrowded, and the train traffic is immediately regulated according to the new headway time. As a result, the number of passengers on the platform and their total waiting time can be significantly reduced. Numerical results are provided to show the effectiveness of the proposed method in improving the operation performance, while minimizing the passengers' discomfort.
\end{abstract}

\section{INTRODUCTION}

Nowadays, metro lines are experiencing an exponential growth of transportation demand, which often leads to low performance of the transport service. In particular, when a temporary large increase of passenger flow occurs, it can provoke train overcrowding, stranded passengers, and delays. With the aim of alleviating passengers discomfort and minimizing delays in the metro lines, traffic regulation actions must be frequently adopted  \cite{Corman2017}, thus resulting in the resolution of the so-called metro traffic regulation problems.

Generally, according to their objectives, metro traffic regulation problems can be classified into two categories: operation-oriented optimization and passenger-oriented optimization. The former focuses on the minimization of operation costs, e.g., trains' delays and energy consumption; while the latter aims at improving service quality, e.g., reducing  the average passenger waiting time and  the severity of trains' and stations' crowdedness. Only recently, a few works on metro traffic regulation have integrated operation-oriented optimization with passenger-oriented optimization, with the aim of considering the influence of train regulation on the service quality for passengers. In particular, it has been proven that, if the passenger travel demand remains constant over a time period, the service quality, in terms of average waiting time and passengers' overcrowding in stations and on-board, can be improved by operation-oriented adjustments, i.e., control actions that only aim at maintaining punctuality and nominal headway times \cite{Moaveni2018,Li2019}. However, in reality, the travel demands of passengers vary with time. As a result, under a temporarily large passenger flow, the above regulation methods may not be effective. 


 In the related literature, several works address passenger-oriented optimization. To increase the service quality in presence of time-varying passenger demands, Wang et al. \cite{Wang2015} investigate the generation of schedules with non-fixed headway, which depends on the number of waiting passengers and the passenger arrival rate at stations. Authors prove that the performance of the non-fixed headway train schedule is better than the one of traditional fixed-headway train schedule. Taking the train capacity into consideration, Zhang et al. \cite{Zhang2018} focus on a comprehensive timetable optimization problem that aims at minimizing the passenger total travel time while keeping the train load factor at a reasonable level. The obtained timetable shows that, increasing the frequency of trains, passengers' congestion can be avoided and the total travel time can also be reduced. The results in \cite{Li2021} prove that a small level of flexible headway  results in better performance in regenerative energy absorption. Finally, Cavone et al. \cite{Cavone2020} develop a technique to reschedule the traffic to minimize both the train delays and the number of passengers suffering from the side-effects caused by rescheduling. Note that for all the discussed works the objective of traffic regulation is the optimization of a trade off between both the operation and service quality, by applying operational control actions.\par

In addition, the following works combine operation-oriented optimization with passenger-oriented optimization. In  \cite{Shi2018}, an integrated integer linear programming problem is formulated, where the control actions regard both the timetable rescheduling and the passenger flow control. By solving the problem, the total waiting time of passengers is minimized, and the optimal passenger flow control policy is obtained. In \cite{Besinovic2019}, an iterative train and passenger management framework  is proposed. Faced with disruptions, the train traffic management model incorporates the passenger flow management model to automatically reschedule trains and control passenger flows. As a result, the total delays of passengers, number of denied passengers, and adjustments of trains are minimized.\par

Differently from the aforementioned works, this paper addresses the real-time metro traffic regulation problem under disturbances while integrating the real-time headway optimization to improve the flexibility and capability of transporting passengers. The proposed approach considers two modules: the train operation module (TOM) and the passenger flow module (PFM). The TOM is in charge of traffic control to improve the operation performance under disturbances, and the PFM is in charge of monitoring the number of passengers and calculating an optimal headway when passengers' congestion happens. Hence, the objective of the TOM is to optimize the trade off of punctuality, regularity, and control effort by means of control actions that consist in adjusting the running time and the dwell time of trains. Differently, the objective of the PFM is to optimize the trade off between the minimization of the passengers waiting time and the maximization of the load rate of trains, by applying a non-fixed headway policy. The optimized headway obtained with the PFM module is transmitted to the TOM to control the trains. As a result, the density of trains is increased, and the crowding level on the platforms is relieved. The TOM remains continuously active to manage traffic delays, while the PFM is activated only when overcrowding at stations occurs. The interaction between the two modules is iterated until the overcrowding is reduced below the maximum platforms capacity. Such a service-oriented metro traffic regulation method makes the rescheduled timetable consistent with time-dependent passenger travel demands. The proposed method is tested on the Beijing metro line 9 via simulations, showing that the number of passengers on platforms and their waiting time can be quickly and significantly reduced with respect to the case of traffic regulation based on fixed headway.

The remainder of this paper is organised as follows. Section \ref{Problem description} describes the problem considered in this work.  Section \ref{Formulation} presents the mathematical formulations of the passenger flow and the traffic operation. Section \ref{Methodology}  explains in detail the proposed service-oriented traffic regulation (STR) framework. In Section \ref{case study}, a case study is presented to demonstrate the effectiveness of the method. Finally, conclusions and future developments are summarized in Section \ref{conclusions}.

\section{PROBLEM DESCRIPTION}\label{Problem description}

\begin{figure}[!tb]
  \centering\includegraphics[width=\hsize]{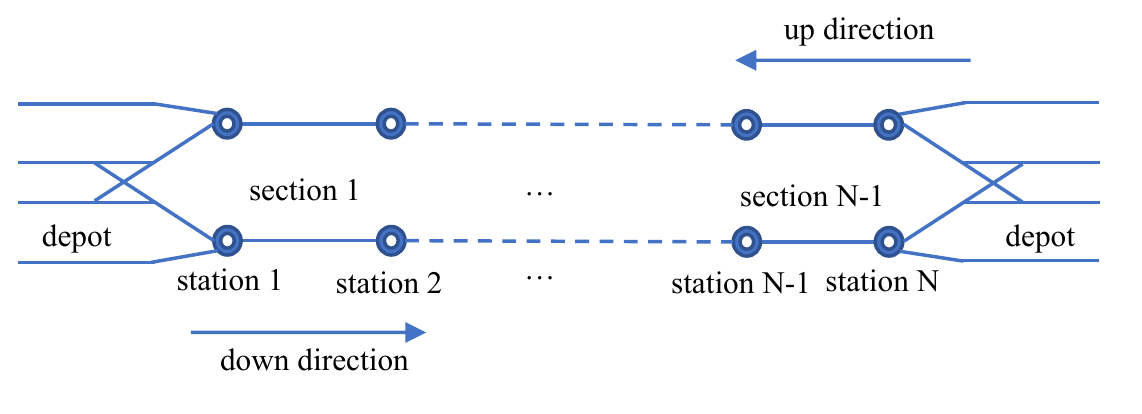}
  \caption{Layout of a closed loop metro line.}
  \label{fig.layout}
\end{figure}

The layout of a generic closed loop metro line is shown Fig.~\ref{fig.layout}, in which the two directions (up direction and down direction) are regarded as two separate lines where trains are operated separately. In this work we only consider a single direction metro line which contains N stations denoted by the set $\textbf{J}=\{j|j=1,2,\dots,\text{N}\}$, two depots located at the terminal along the line (station $1$ and station N), and trains circulating on the line denoted by the set $\textbf{I}=\{i|i=1,2,\dots,\text{M}\}$. Note that, neither train overtaking and crossing are normally allowed, nor the skip-stop pattern.\par
In practical operations, due to the occurrence of unpredicted events, such as equipment failures or hindrance of doors by passengers, slight delays occur from time to time. Therefore, metro traffic regulation is to compensate the time deviations between actual and planned operations. Furthermore, during peak hours, the dramatic increase of arriving passengers may lead to crowded trains, or, in the worst case, when the maximal train capacity is exceeded, some waiting passengers may be prevented  from boarding the first coming train and left on the platform. In this condition, with the accumulation of arriving passengers, the total number of waiting passengers can temporally exceed the acceptable maximal capacity of the platform. The adjustment of headway time can be considered to effectively reduce the severity of crowding and thus limit the passengers' total waiting time.

\begin{figure}[!tb]
  \centering\includegraphics[width=\hsize]{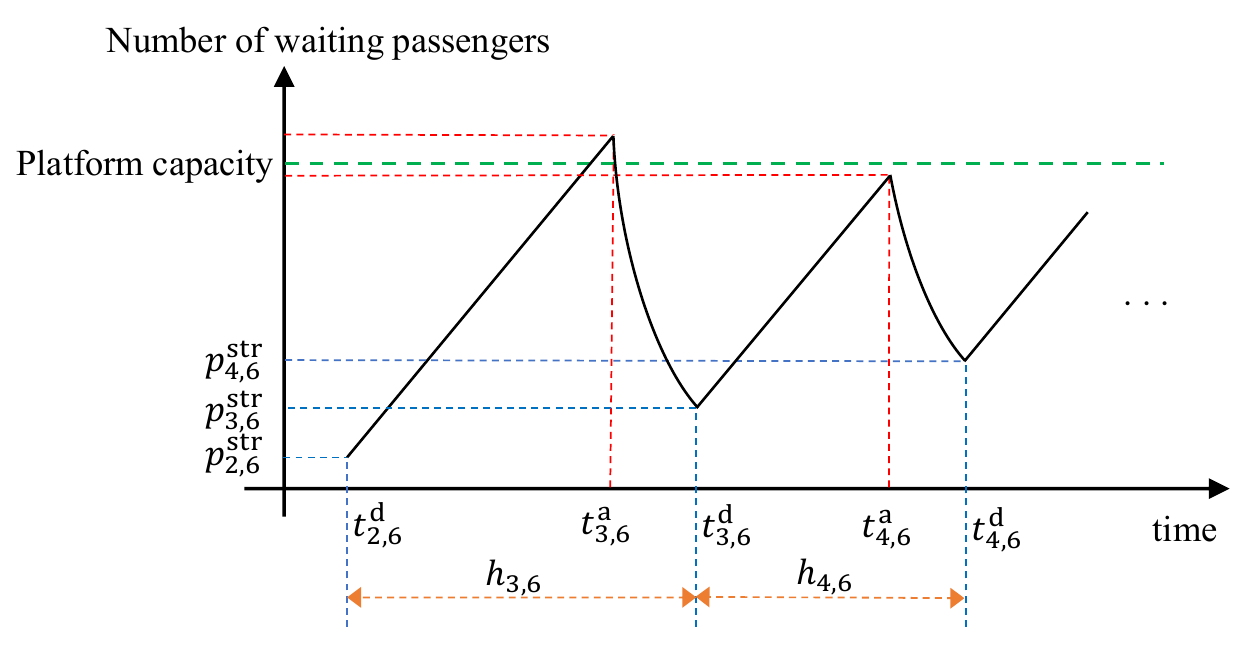}
  \caption{Number of waiting passengers with non-fixed headway.}
  \label{fig.example}
\end{figure}

In this paper a service-oriented regulation method is presented that aims at minimizing the passengers' waiting time by optimally rescheduling the trains' headway time in case of delays and large increased passenger flow in a metro line.  Figure~\ref{fig.example} presents an example to illustrate the characteristics of the proposed service-oriented regulation method. In this example, the number of waiting passengers at station $6$ is represented by a black line. When train $2$ departs from station $6$, some passengers denoted by $p^{\text{str}}_{2,6}$ are left on the platform in relation to the limit of train capacity. The headway time between train $2$ and the successive train $3$ is the nominal one, denoted by $h_{3,6}$. Before the arrival of the successive train $3$, the number of passengers waiting at station $6$ increases according to the number of arriving  passengers at station $6$,  and exceeds the platform capacity denoted by a green dashed line. If the headway time  $h_{4,6}$ is properly reduced with respect to $h_{3,6}$, then the number of waiting passengers can be reduced and consequently kept lower than the platform capacity. This example shows that, if the passenger arrival rate is greater than expected, the proper modification of headway times between trains is an effective way to maintain the number of waiting passengers below the platform capacity and consequently reduce the passenger waiting time.\par

\section{PROBLEM FORMULATION}\label{Formulation}

This section first provides a mathematical model of the passenger flow in a metro line, then a mathematical model of train operations, both represented as event-driven models, in which the status of the system evolves when a departure of a train from a station occurs in the line. The definitions of the variables and parameters of the passenger flow model are listed in Table \ref{Tab:defination on passenger}, while the ones of the trains operation model are reported in Table \ref{Tab:defination on train}.
%
%
%
\subsection{Passenger Flow Model}

In daily operations, metro trains stop at each station for passengers to alight and/or board. The number of on-board passengers when train $i$ departs from station $j$ is represented by $p^{\text{in}}_{i,j}$ and can be calculated as
\begin{equation}
    p^{\text{in}}_{i,j}=p^{\text{in}}_{i,j-1}+p^{\text{boa}}_{i,j}-p^{\text{ali}}_{i,j}, \hspace{1cm} \forall i, \forall j,\label{loading passengers}
\end{equation}
where $p^{\text{boa}}_{\text{i},j}$ is the number of passengers boarding on train $i$ at station $j$ and $p^{\text{ali}}_{i,j}$ is the number of passengers alighting from train $i$ with destination station $j$. Equation~\eqref{loading passengers} describes the number of on-board passengers during the running of train $i$. Furthermore, the number of alighting passengers from train $i$ at station $j$ is proportional to the number of on-board passengers departing from the last station $j-1$ as follows:
\begin{equation}
    p^{\text{ali}}_{i,j}=\beta_{i,j} p^{\text{in}}_{i,j-1}, \hspace{1cm} \forall i, \forall j,\label{alighting passengers}
\end{equation}
where $\beta_{i,j}$ is a proportional factor which can be estimated from the OD (origin-destination) matrix.\par
Although the features of arriving passengers at a station are usually time varying, it is still reasonable that the number of passengers arriving at the station is distributed uniformly over the headway time \cite{Hou2019}. Thus, the number of arriving passengers  for train $i$ during the period of a headway, denoted by $p^{\text{arr}}_{i,j}$, is given as
\begin{equation}
    p^{\text{arr}}_{i,j}=\alpha_{i,j} h_{i,j}, \hspace{1cm} \forall i, \forall j, \label{arriving passengers} 
\end{equation}
in which $\alpha_{i,j}$ is the rate reflecting the number of arriving passengers at station $j$ per second  before train $i$ leaves station $j$, $h_{i,j}$ represents the headway time between two successive trains $i$ and $i-1$.\par
Due to the limit on train capacity, when train $i$ dwells at station $j$, not all the waiting passengers can board on the train. The number of passengers allowed to board the train depends on the remaining train capacity, which is related to  the number of boarding and alighting passengers at the preceding stations. 
Then the number of passengers boarding on train $i$ at station $j$ can be expressed as
\begin{equation}
    p^{\text{boa}}_{i,j}=\text{min}\{( P^{\text{cap}}-p^{\text{in}}_{i,j-1}+p^{\text{ali}}_{i,j}),(p^{\text{arr}}_{i,j}+p^{\text{str}}_{i-1,j})\}, \hspace{0.5cm} \forall i, \forall j,\label{boarding passengers}
\end{equation}
where $P^{\text{cap}}$ is the capacity of one train, and $p^{\text{str}}_{i-1,j}$ represents the stranded passengers who can not board on the front train $i-1$. \par

The number of stranded passengers left on the platform when train $i$ departs from station $j$ can thus be written as
\begin{equation}
    p^{\text{str}}_{i,j}=p^{\text{str}}_{i-1,j}+p^{\text{arr}}_{i,j}-p^{\text{boa}}_{i,j}, \hspace{1cm} \forall i, \forall j.\label{stranded passengers}
\end{equation}
Substituting \eqref{alighting passengers}, \eqref{arriving passengers}, and \eqref{boarding passengers} into \eqref{loading passengers}, the number of on-board passengers can be further calculated as
\begin{equation}
    p^{\text{in}}_{i,j}=\text{min}\{ P^{\text{cap}},p^{\text{str}}_{i-1,j}+\alpha_{i,j} h_{i,j}+(1-\beta_{i,j}) p^{\text{in}}_{i,j-1}\},  \forall i, \forall j.
    \label{Dynamic: on-board passengers}
\end{equation}
Moreover, by substituting \eqref{alighting passengers},  \eqref{arriving passengers} and \eqref{boarding passengers} into \eqref{stranded passengers}, a nonlinear equation on the number of stranded passengers can be expressed as
 \begin{equation}
 \begin{split}
    p^{\text{str}}_{i,j}=\text{max}\{0,p^{\text{str}}_{i-1,j}&+\alpha_{i,j} h_{i,j}\\+&(1-\beta_{i,j}) p^{\text{in}}_{i,j-1}- P^{\text{cap}}\},  \forall i, \forall j.
    \end{split}
    \label{Dynamic: stranded passengers}
\end{equation}

\begin{figure}[tb]
  \centering\includegraphics[width=\hsize]{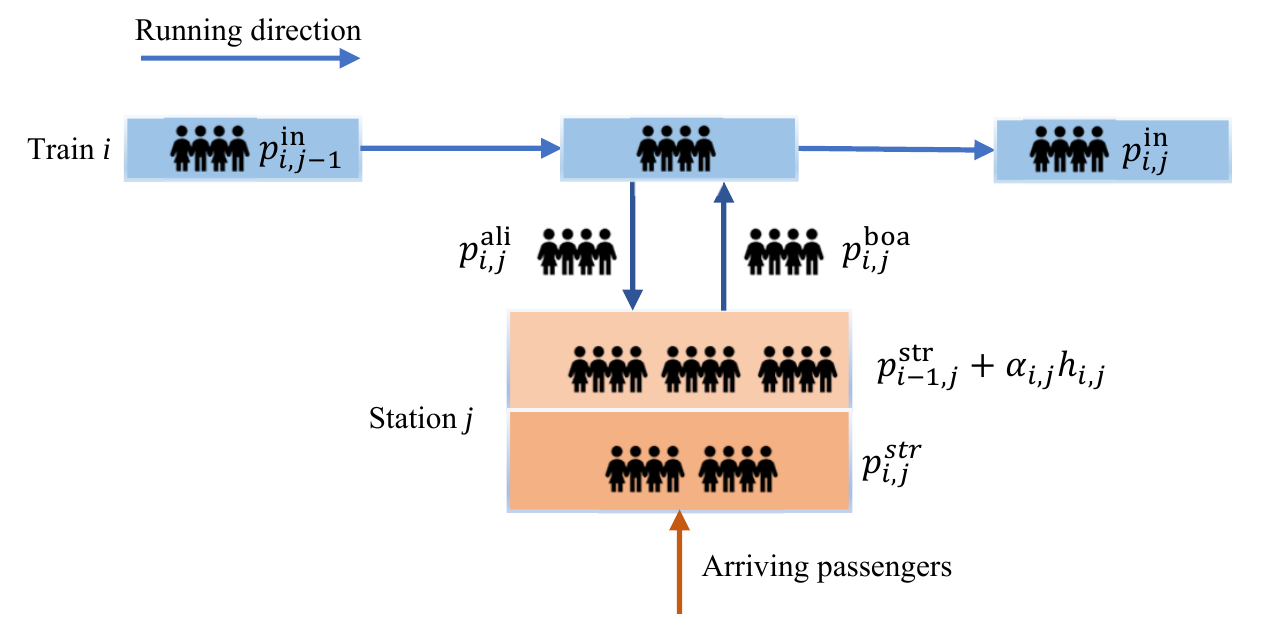}
  \caption{Illustration of passenger dynamics.}
  \label{fig.passenger dynamics}
\end{figure}

Generally, a congestion at a platform may occur when passengers  are boarding or alighting the train, then we compute the maximal number of passengers when train $i$ dwells in station $j$ as $p_{i,j}$ as below
\begin{equation}
    p_{i,j}=p^{\text{str}}_{i-1,j}+\alpha_{i,j}h_{i,j}+\beta_{i,j}p^{\text{in}}_{i,j-1}, \hspace{1cm}\forall i, \forall j.
\end{equation}
 Since the over-crowded passenger flow usually results in higher risk disastrous consequences (e.g., stampedes),  it is necessary to keep the  number of passengers on the platform below an acceptable level. Therefore, measures should be taken to erase the heavy congestion, also reducing the waiting time of passengers at the station. An effective way to avoid over-crowding at platforms is to improve the traffic density, that is by considering non-fixed headway times between successive trains.  Figure~\ref{fig.passenger dynamics} illustrates the quantitative relation of passengers' flow inside of a train $i$ and at a platform $j$.  When train $i$ arrives at station $j$, it carries $p_{i,j-1}^\text{in}$ passengers and there are $p_{i,j}^\text{ali}$ passengers alighting from the train. Before its departure, there are $p^{\text{str}}_{i-1,j}$ stranded passengers of train $i-1$ plus $\alpha_{i,j}h_{i,j}$ new arriving passengers waiting at the platform to board on train $i$. However, due to the limit of train capacity, only $p_{i,j}^\text{boa}$ passengers can board train $i$ successfully. After train $i$ departures from station $j$, it transports $p_{i,j}^\text{in}$ passengers, while $p_{i,j}^\text{str}$ passengers are left at the platform.\par

 Notice that the number of on-board passengers and the number of stranded passenger at a station will be up-dated after the train leaves the station. Namely, the passenger flow model evolves in accordance with the departure of a train from a station.

In this paper, we propose an event-triggered optimizing procedure to obtain an optimal headway, and its enabling condition is defined as
\begin{equation}\label{eq:cond}
p_{i,j} \geq  P^{\text{acc}}, \, \text{if } \; \exists \;  i \in \textbf{I}, \; \text{and} \ j \in \textbf{J},
\end{equation}

\noindent where $P^\text{acc}$ is the maximal number of passengers that can be accommodated safely at one platform.  Suppose $p_{i,j}\geq P^{\text{acc}}$ when train $i$ is at station $j$. Then the optimization problem, formulated in~\eqref{eq:f}, will be triggered as soon as train $i$ leaves station $j$. The objective of the procedure is to calculate an optimal headway $\hat h$, which is the same for all the trains on the line, so as to optimize the trade-off between the weighted sum of the total passenger waiting time for the next coming trains $i+1$ and their total load rates along the metro line.


\begin{equation}\label{eq:f}
    \text{min}\;F= \omega_\text{w}\frac{\sum\limits^{\text{N}}_{j=1}T^{\text{wait}}_{i+1,j}}{\sum\limits^{\text{N}}_{j=1}T^{\text{wait}\prime}_{i+1,j}}- \omega_\text{l}\frac{\sum\limits^{\text{N}}_{j=1}L^{\text{load}}_{i+1,j}}{\sum\limits^{\text{N}}_{j=1}L^{\text{load}\prime}_{i+1,j}}, \text {for} \; i=\{1, \dots, M\},
\end{equation}
subject to:
\begin{equation}
\begin {split}
    p^{\text{str}}_{i+1,j}=\text{max}\{0,p^{\text{str}}_{i,j}&+\alpha_{i+1,j}\hat h+\\&(1-\beta_{i+1,j}) p^{\text{in}}_{i+1,j-1}-P^{\text{cap}})\}, \; \forall i, \forall j,
    \end {split}
\end{equation}
\begin{equation}
\begin{split}
        p^{\text{in}}_{i+1,j}=\text{min}\{P^{\text{cap}},p^{\text{str}}_{i,j}+&\alpha_{i+1,j}\hat h+\\&(1-\beta_{i+1,j}) p^{\text{in}}_{i+1,j-1}\}, \; \forall i, \forall j,
        \end{split}
\end{equation}
\begin{equation}\label{eq:T_wait}
 T^{\text{wait}}_{i+1,j}=p^{\text{str}}_{i,j}\hat h+\frac{1}{2}\alpha_{i+1,j}{\hat h}^2,  \hspace{1cm} \forall i, \forall j,
\end{equation}
\begin{equation}\label{eq:L_load}
    L^{\text{load}}_{i+1,j}=p^{\text{in}}_{i+1,j}/P^{\text{cap}}, \hspace{1cm} \forall i, \forall j,
\end{equation}
\begin{equation}\label{eq:min_H}
    \hat h\geq H_{\text{min}},
\end{equation}

\noindent where $\omega_\text{w}$ and $\omega_{\text{l}}$ are weight factors  that reflect the trade-off between reducing the total passenger waiting time and improving the carrying capacity,  and $H_{\text{min}}$ is the minimal allowable headway for safety constraint. {The decision variable $\hat h$ is the headway between two successive trains $i$ and $i+1$ at station $j$. $T^{\text{wait}}_{i+1,j}$ is the total waiting time of stranded passengers left by train $i$ and new arriving passengers before the departure of train $i+1$. As shown in Fig.~\ref{fig.eg_waitingtime}, $T^{\text{wait}}_{i+1,j}$ equals the size of the shadowed area.
In~\eqref{eq:L_load}, $L^{\text{load}}_{i+1,j}$ is the load rate of train $i+1$ at station $j$. In order to transform the two indicators in \eqref{eq:T_wait} and \eqref{eq:L_load} to the same magnitude, in~\eqref{eq:f}, they are divided by $T^{\text{wait}\prime}_{i+1,j}$ and $L^{\text{load}\prime}_{i+1,j}$, respectively, which are two nominal values of passenger waiting time and load rate with the current headway. Note that the performance index $f$ for passenger service is defined on the line level rather than  an individual station, thus it  also reflects the system's performance in the near future.  In summary, whenever the number of passengers on a platform exceeds the maximum acceptable threshold, the headway will be recalculated such that the performance index $f$ defined in~\eqref{eq:f} is optimized.

\begin{figure}[!htb]
  \centering\includegraphics[width=0.8\hsize]{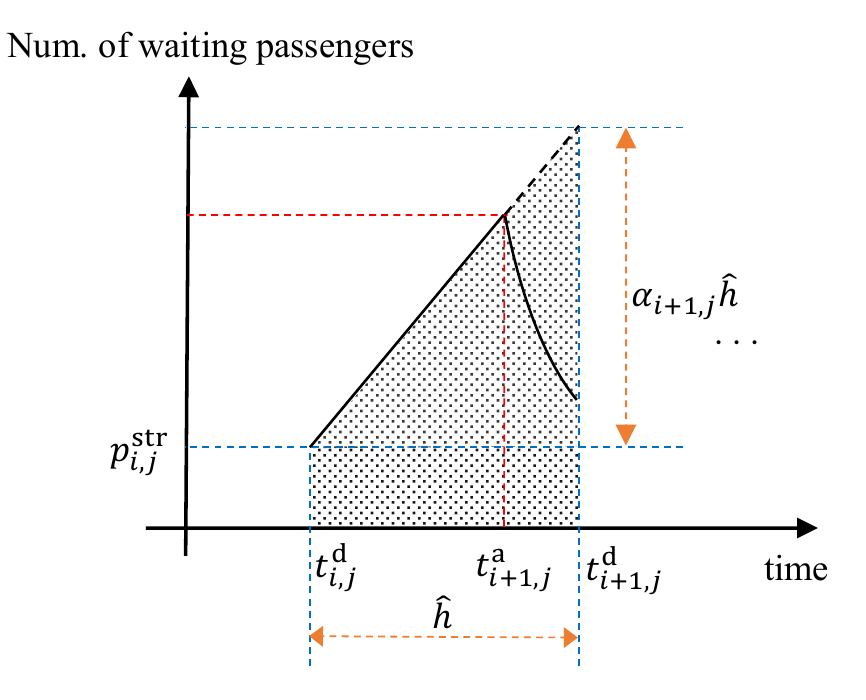}
  \caption{Passenger waiting time.}
  \label{fig.eg_waitingtime}
\end{figure}


\begin{table*}[!htb]
  \centering
  \caption{Variables and Parameters of Passenger Flow}
  \label{Tab:defination on passenger}
  \begin{tabular}{|c||c|}
    \hline
    Notation         & Definition \\ \hline
    $p^{\text{in}}_{i,j}$         & Number of on-board passengers when train $i$ departs from station $j$ \\ \hline
    $p^{\text{boa}}_{i,j}$       & Number of passengers boarding on train $i$ at station $j$ \\ \hline
    $p^{\text{ali}}_{i,j}$     & Number of passengers alighting from train $i$ with destination station $j$ \\ \hline
    $p^{\text{arr}}_{i,j}$      & Number of passengers arriving at station $j$ during  the headway period between train $i$ and train $i-1$\\ \hline

    $p^{\text{str}}_{i,j}$       &  Number of stranded  passengers  left  at  the  platform   after train $i$ departs from station $j$  \\ \hline
    $p_{i,j}$       &  Maximal number of passengers  when train $i$ dwells at station $j$  \\ \hline
    $P^{\text{cap}}$       &   Capacity of one train  \\ \hline
    $P^{\text{acc}}$       &  Maximum acceptable capacity of one platform  \\ \hline
    $\alpha_{i,j}$       &  Arrival rate of passengers at station $j$ per second waiting for train $i$\\ \hline
    $\beta_{i,j}$          &   Proportional factor of passengers alighting from train $i$ at station $j$\\ \hline
    $h_{i,j}$   & Actual headway between train $i$ and train $i-1$ at station $j$\\ \hline
    $ \hat h$ & Optimal headway between trains\\\hline
    $H_{\text{min}}$ & Minimal allowable headway for safety\\\hline
  \end{tabular}
\end{table*}

\begin{table*}[!htb]
  \centering
  \caption{Variables and Parameters of Train Operation}
  \label{Tab:defination on train}
  \begin{tabular}{|c||c|}
    \hline
    Notation         & Definition \\ \hline
    $t^{\text{d}}_{i,j}$ & Actual departure time of train $i$ at station $j$ \\
    \hline
    $\text{T}_{i,j}$ & Scheduled departure time of train $i$ at station $j$\\
    \hline
    $r_{i,j}$ & Actual running time of train $i$ at section $j$ \\
    \hline
    $s_{i,j+1}$ & Actual dwell time of train $i$ at station $j+1$ \\
    \hline
    $R_j$ & Nominal running time at section $j$ \\
    \hline
    $D$ & Minimal dwell time when no passengers get on/off \\
    \hline
    $\lambda_{i,j+1}$ & Delay rate of station $j$ when train $i$ dwells \\
    \hline
    $u^{\text{1}}_{i,j}$ & Control strategy implemented to adjust the running time of train $i$ at section $j$ \\\hline
    $u^{\text{2}}_{i,j}$ & Control strategy implemented to adjust the dwell time of train $i$ at station $j$ \\\hline
    $w^{\text{1}}_{i,j}$ & Effect of external disturbance to the running process of train $i$ at section $j$ \\\hline
    $w^{\text{2}}_{i,j}$ & Effect of external disturbance to the dwell process of train $i$ at station $j$ \\\hline
    $U^{\text{1}}_{\text{min}}$ & Minimal allowable control adjustment on running time \\\hline
    $ U^{\text{2}}_{\text{min}}$ & Minimal allowable control adjustment on dwell time \\\hline
    $U^{\text{1}}_{\text{max}}$ & Maximal allowable control adjustment on running time \\\hline
    $U^{\text{2}}_{\text{max}}$ & Maximal allowable control adjustment on dwell time \\\hline
  \end{tabular}
\end{table*}

\subsection{Trains Operation Model}

For metro systems, the train overtaking and crossing operations are prohibited,  and usually} trains are operated with fixed all-stop patterns from the first station to the last station.  Moreover, the running time of the train can not be changed during the running operation in sections. Therefore, any control strategy should be decided before the train departs from a station.
Based on the layout in Fig.~\ref{fig.layout}, the traffic operation process can be described in a macroscopic view that only arrival/departure times are adjusted under practical constraints.
The departure time constraint is formulated as follows: 
\begin{equation}
    t^{\text{d}}_{i,j+1}=t^{\text{d}}_{i,j}+r_{i,j}+s_{i,j+1}, \hspace{1cm} \forall i, \forall j,\label{departure time}
\end{equation}
where $t^{\text{d}}_{i,j}$ denotes the actual departure time of train $i$ at station $j$, $r_{i,j}$ is the actual running time from stations $j$ to $j+1$, $s_{i,j+1}$ is the actual dwell time at station $j+1$. Specifically, the running time $r_{i,j}$ can be further expressed as
\begin{equation}
    r_{i,j}=R_{j}+u^{\text{1}}_{i,j}+w^{\text{1}}_{i,j}, \hspace{1cm} \forall i, \forall j,\label{running time}
\end{equation}
where $R_j$ is the nominal running time at section $j$, $u^{\text{1}}_{i,j}$ is the control strategy to adjust the running time of train $i$ at section $j$, and $w^{\text{1}}_{i,j}$ is the external disturbance to the running time.
The actual dwell time of train $i$ at station $j+1$ is
\begin{equation}
    s_{i,j+1}=D+\lambda_{i,j+1}(t^{\text{d}}_{i,j+1}-t^{\text{d}}_{i-1,j+1})+ u^{\text{2}}_{i,j+1}+w^{\text{2}}_{i,j+1},  \forall i, \forall j, \label{dwell time}
\end{equation}
where $\lambda_{i,j+1}$ is the delay rate, $D$ is the minimal dwell time when no passenger gets on the train and the doors are closed as soon as possible, $u^{\text{2}}_{i,j+1}$ is the control strategy to adjust the dwell time of train $i$ at station $j+1$,  and $w^{\text{2}}_{i,j+1}$ is the external disturbance to the dwell time that is mainly caused by the hinder of the closing process of the platform screen doors especially in an over-crowded situation.\par
\textbf{Remark:} The delay rate reflects the effect of headways between successive trains to the dwell time for boarding passengers.  According to the estimation method presented in \cite{Moaveni2018}, the delay rate $\lambda_{i,j+1}$ is related to the headway and the entering passenger flow. Note that the number of passengers on the platform is bounded, therefore if the passenger flow exceeds capacity to board, $\lambda_{i,j+1}$ will be  considered saturated.

By substituting~\eqref{running time} and~\eqref{dwell time} into \eqref{departure time}, the departure time of train $i$ can be described as
\begin{equation}
\begin{split}
t^{\text{d}}_{i,j+1}=t^{\text{d}}_{i,j}+\lambda_{i,j+1} (t^{\text{d}}_{i,j+1}&-t^{\text{d}}_{i-1,j+1})+R_{j}+\\&D+u_{i,j}+w_{i,j},  \forall i, \forall j,\label{dynamics}
\end{split}
\end{equation}
where
\begin{equation}
u_{i,j}=u^{\text{1}}_{i,j}+u^{\text{2}}_{i,j+1}, \hspace{1cm}\forall i, \forall j,\label{u}
\end{equation}
\begin{equation}
w_{i,j}=w^{\text{1}}_{i,j}+w^{\text{2}}_{i,j+1}, \hspace{1cm}\forall i, \forall j.\label{w}
\end{equation}\par
According to the nominal timetable, a scheduled train operation model can be constructed
\begin{equation}
\begin{split}
      T^{\text{d}}_{i,j+1}=T^{\text{d}}_{i,j}+\lambda_{i,j+1} (T^{\text{d}}_{i,j+1}-&T^{\text{d}}_{i-1,j+1})+\\&R_{j}+D, \forall i, \forall j.
\end{split}\label{schedule}
\end{equation}
 Note that trains are supposed to run regularly, i.e., the headway $T^{\text d}_{i+1,j}-T^{\text d}_{i,j}=H$ is scheduled as constant for all $i\in \textbf{I}$ and $j\in \textbf{J}$. We define the time deviation as $x^{\text{d}}_{i,j+1}=t^{\text{d}}_{i,j+1}-T^{\text{d}}_{i,j+1}$. By subtracting \eqref{schedule} from \eqref{dynamics}, the time deviation of the train operation model can be obtained
\begin{equation}
   x^{\text{d}}_{i,j+1}=x^{\text{d}}_{i,j}+\lambda_{i,j+1} (x^{\text{d}}_{i,j+1}-x^{\text{d}}_{i-1,j+1})+u_{i,j}+w_{i,j}, \forall i, \forall j.\label{transition}
\end{equation}
 Since whenever the number of passengers $p_{i,j}$ at any platform is larger than the platform capacity, the optimizing procedure on headway is triggered, a new scheduled timetable will be generated according to the obtained optimal headway  $\hat h$. In this case, the time deviation is replaced by
\begin{equation}
     x^{\text{d}}_{i,j}= t^{\text d}_{i,j}-T^{\text d}_{i,j}+{\hat h}'-\hat h, \forall i, \forall j, 
     \label{if headway}
\end{equation}
where $\hat h'$ is the previous scheduled headway, and $\hat h$ is the new one obtained by solving problem \eqref{eq:f}-\eqref{eq:min_H}.

 For the train operation process, we  also consider the following safety and control constraints:
\begin{itemize}
    \item To keep the safety distance of successive trains tracking interval, we have the following constraint on the departure headway:
    \begin{equation}
         x^{\text{d}}_{i,j+1}-x^{\text{d}}_{i-1,j+1}\geq H_{\text{min}}-\hat h, \forall i, \forall j.\label{minimal headway}
    \end{equation}
    \item Constraints on the control inputs:
    \begin{equation}
         U^{\text{1}}_{\text{min}} \leq u^{\text{1}}_{i,j} \leq U^{\text{1}}_{\text{max}},
        U^{\text{2}}_{\text{min}} \leq u^{\text{2}}_{i,j+1} \leq U^{\text{2}}_{\text{max}},  \forall i, \forall j. \label{control constraints}
    \end{equation}
\end{itemize}

The metro traffic regulation problem is then formulated as follows:
\begin{equation}\label{eq:J}
\begin {split}
    \text{min}\;J=\sum_{i\in I}\sum_{j\in J}a_{\text{t}}(x^{\text{d}}_{i,j})^2&+b_{\text{t}}(x^{\text{d}}_{i,j}-x^{\text{d}}_{i-1,j})^2\\&+c_{\text{t}}((u^1_{i,j})^2+(u^2_{i,j})^2),\\
    \text{s.t.}\; \eqref{transition}, \eqref{if headway}, \eqref{minimal headway}, \eqref{control constraints}. \\
\end {split}
\end{equation}
where $a_{\text{t}}$, $b_{\text{t}}$, and $c_{\text{t}}$ are weight factors  that reflect the trade-off between punctuality, regularity, and control effort, respectively.
The resulting optimization problem is a nonlinear quadratic programming model  of the train operation control.


\section{THE PROPOSED SERVICE-ORIENTED REGULATION FRAMEWORK}\label{Methodology}

\begin{figure}[!tb]
  \centering\includegraphics[width=\hsize]{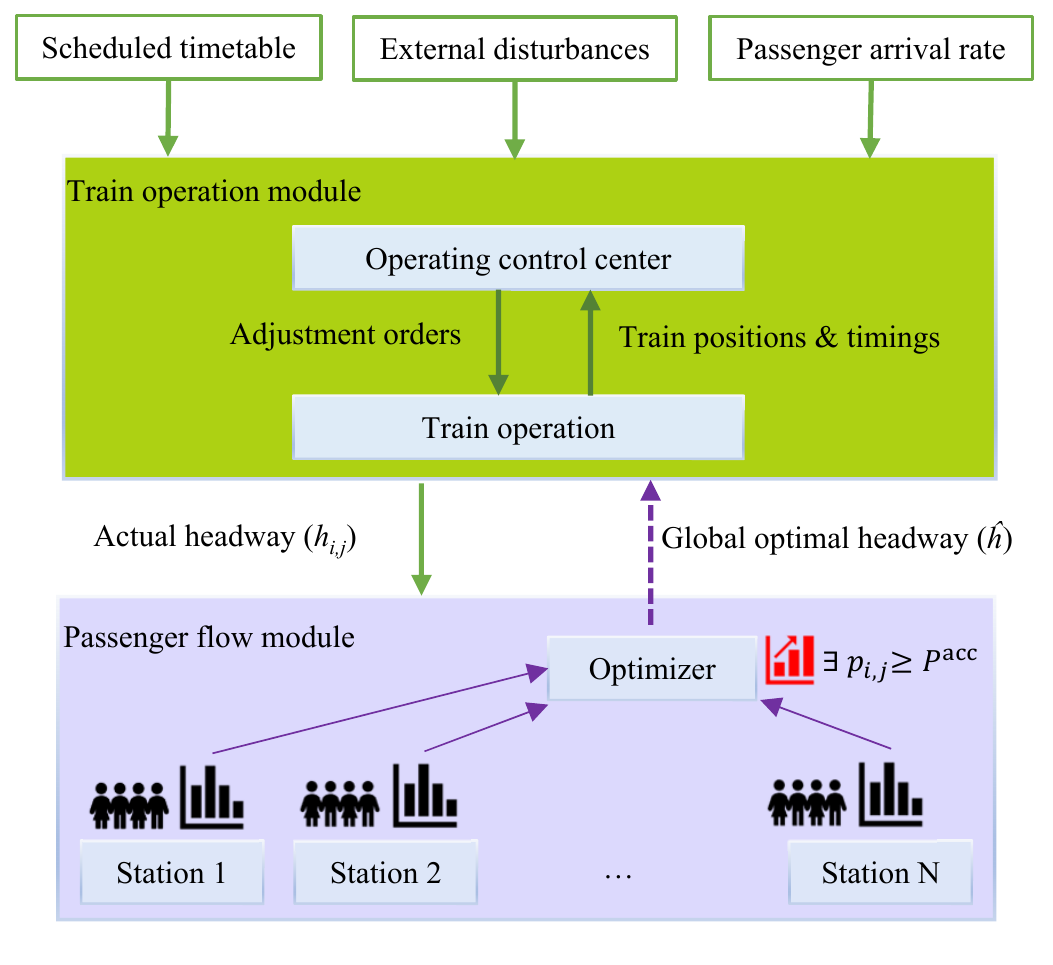}
  \caption{Framework of the service-oriented metro traffic regulation method.}
  \label{fig.framework}
\end{figure}

 In this section, the proposed STR method is presented. The framework of STR is shown in Fig.~\ref{fig.framework},  in which two modules are included:
\begin{itemize}
\item The train operation module (TOM), which computes the operating control actions to adjust running time and dwell time according to the real-time trains' positions and timings.
\item The passenger flow module (PFM), which  computes the optimal headway time in case of overcrowded stations, based on the number of on-board, boarding, alighting, and stranded passengers.
\end{itemize}
Let $H$ be the initially scheduled headway time. The two modules interact with each other in the following way.

As soon as a disturbance occurs in the line, the TOM starts with rescheduling trains' running time and dwell time to minimize the time deviations between actual departure times and scheduled departure times. The optimization problem \eqref{transition}-\eqref{eq:J} is solved and the control actions are computed and applied to the line. In particular, based on the current time deviations $x^{\text{d}}_{i,j}$ of all trains along the metro line, the operating control center provides optimal control inputs, i.e., $u^1_{i,j}$ and $u^2_{i,j+1}$. We highlight that at the first application of the TOM the values of $\hat h'$ and $\hat h$ are both equal to $H$ in constraint \eqref{if headway}.

At each train departure the enabling condition~\eqref{eq:cond} is checked by the PFM and once the  number of the passengers $p_{i,j}$ exceeds the  maximum acceptable platform capacity $P^{\text{acc}}$ in at least one station for at least one train, the  optimization problem   \eqref{eq:f}-\eqref{eq:min_H} is triggered and solved  to have an optimal trade-off between the passengers' waiting time and the load rate of the trains in the line. The obtained optimal headway $\hat h$ is transferred to the TOM for timetable adjustment.
After the next trains' departures, the TOM inputs a new series of actual headway times of trains at stations and the enabling condition~\eqref{eq:cond} in PFM is checked again. The procedure, involving both the TOM and the PFM is then iterated as long as~\eqref{eq:cond} holds.
As soon as $p_{i,j}<P^{\text{acc}}$ for all $i\in \textbf{I}, j\in \textbf{J}$, only the TOM module is active for the management of disturbances and the  headway time can be increased appropriately, even to the original one $H$, for the consideration of energy/cost saving.


\section{CASE STUDY}\label{case study}

\begin{figure}[tb]
  \centering\includegraphics[width=0.75\hsize]{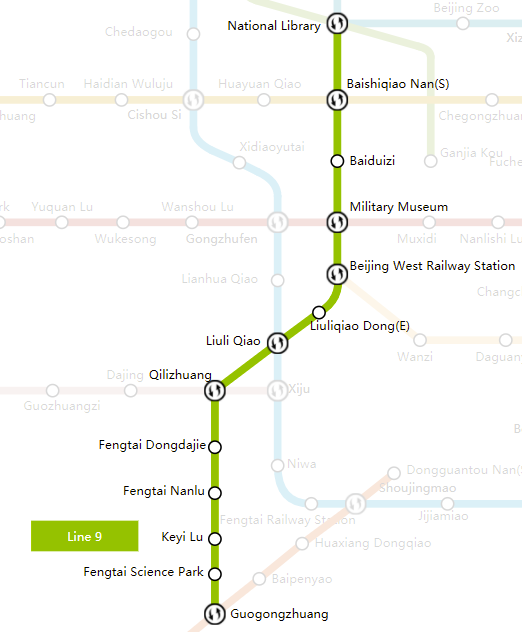}
  \caption{Layout of the Beijing Metro Line 9.}
  \label{fig.line 9}
\end{figure}

We test the  proposed STR method on the case of Beijing Metro line 9 which contains $13$ stations from the station Guogongzhuang to the station National Library (see Fig.~\ref{fig.line 9}). In the case study, the trains that circulate on the line are indexed by $i=\{1, \dots, \text{M}\}$, where M=40. Parameters of train capacity and platform capacity are set as : $P^{\text{cap}}=1860, P^{\text{acc}}=1860$. The weight factors  in \eqref{eq:f} and \eqref{eq:J} are set as: $\omega_\text{w}=0.5, \omega_\text{l}=1.5, a_\text{t}=1,b_\text{t}=1,c_\text{t}=1$. The minimal headway is $H_{\text{min}}=180$s and at the beginning, the scheduled headway $H$ is equal to $360$s. In the train running process, the allowable control action on  the minimal and  the maximal running time are set as $17\%$ and $53\%$ of the nominal running time correspondingly.  The increase or decrease of the adjustable dwell time is not allowed to exceed $10$s. The simulations  are  performed on a $1.60$ GHZ Inter Core i5-10210 CPU,  and the optimization problem are solved by the function fmincon in Matlab, respectively.\par
\begin{figure}[tb]
  \centering\includegraphics[width=0.9\hsize]{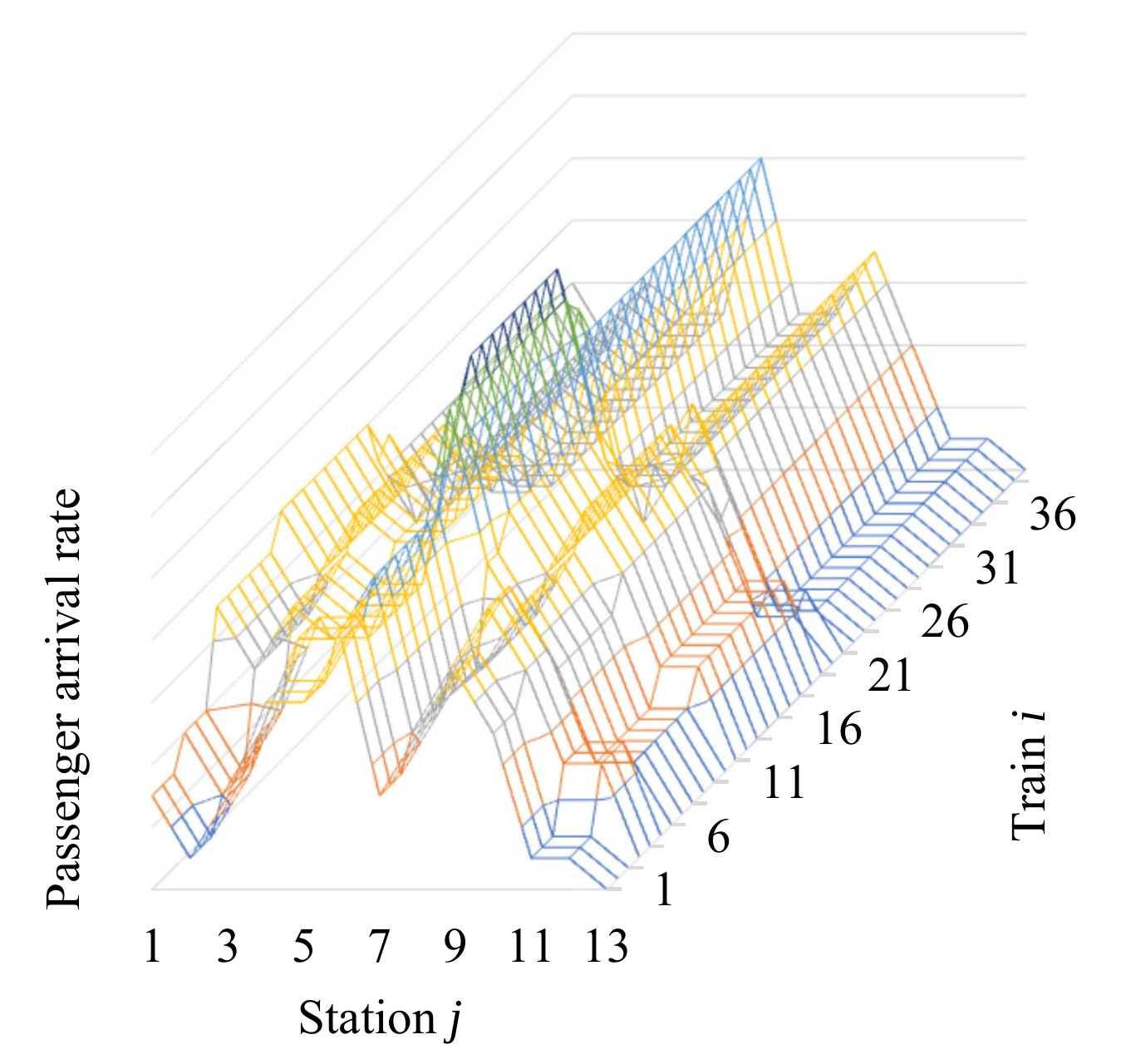}
  \caption{Time-variant parameters of passenger arrival rate.}
  \label{fig.arrival rate}
\end{figure}

Usually the passenger travel demands are time-variant. In Fig.~\ref{fig.arrival rate}, the passenger arrival rate $\alpha_{i,j}$ is plotted.  At some peak hours, the arrival rate of some stations is larger than usual. As a result, the capacity of the train is reached and some waiting passengers are left on the platform. Thus, it is necessary to adjust trains' operation to meet the real-time passenger travel demands. To evaluate the  performance of the proposed regulation method with non-fixed headway in improving operation performance we suppose that an external disturbance occurs in the system on train $13$ at station $1$, and the corresponding variable is set as $w_{13,1}=68$s. Consequently, the TOM module computes the control actions, which are applied to the system. Simultaneously, the number of passengers on platforms, the number of stranded passengers, and the total passenger waiting time are  simulated. Note that, for the sake of simplicity, the evolution of the number of on-board, stranded, and total passengers and the total waiting time are represented in the following figures as function of an event counter $k=i+j$, where $i$ is the index of the train and $j$ is the index of the station.

\begin{figure}[tb]
  \centering\includegraphics[width=\hsize]{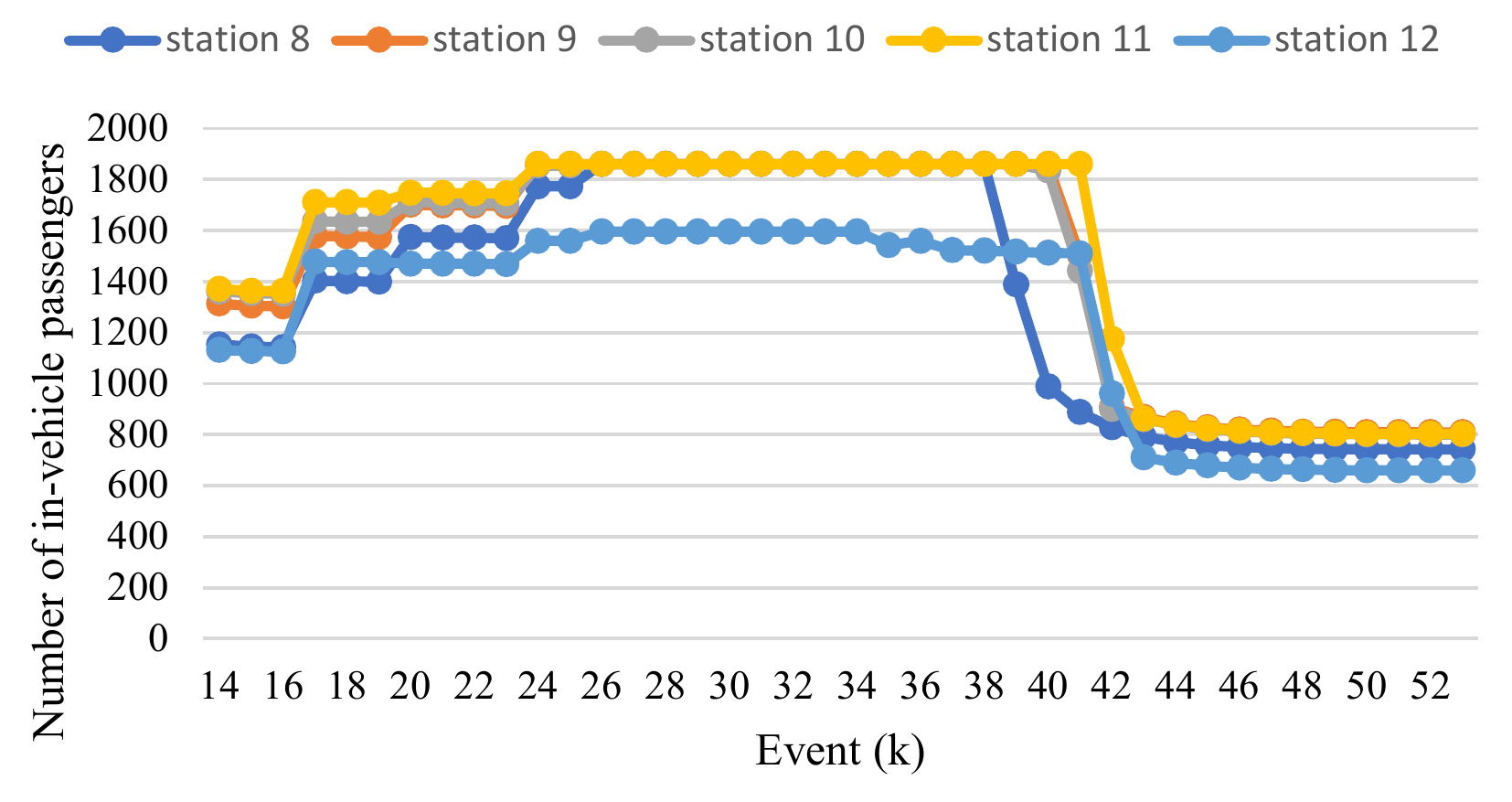}
  \caption{Number of on-board passengers with  the STR method.}
  \label{fig.on-board passengers}
\end{figure}

\begin{figure}[tb]
  \centering\includegraphics[width=\hsize]{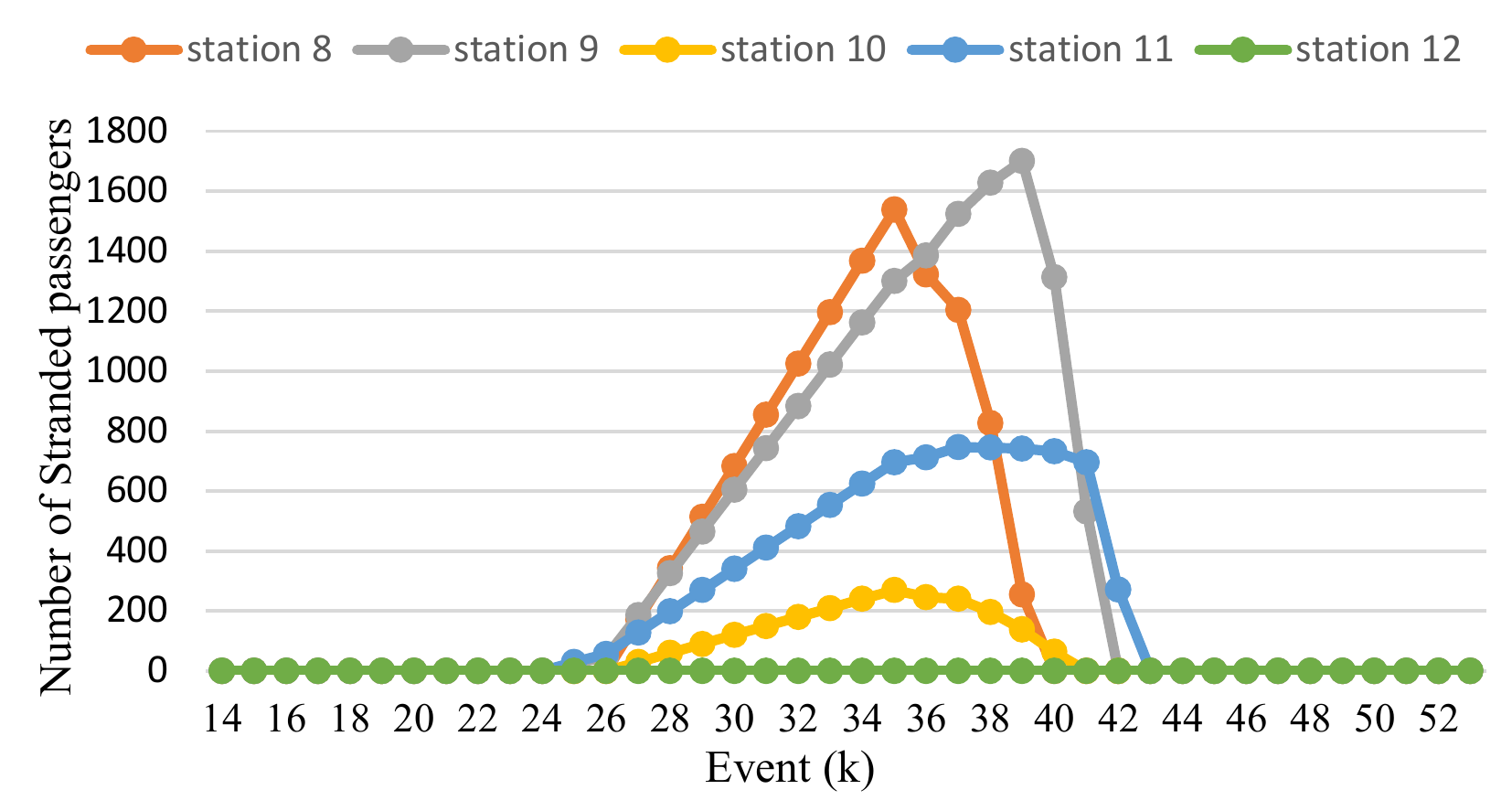}
  \caption{Number of stranded passengers with  the STR method.}
  \label{fig.stranded passengers}
\end{figure}

\begin{figure}[!htb]
  \centering\includegraphics[width=\hsize]{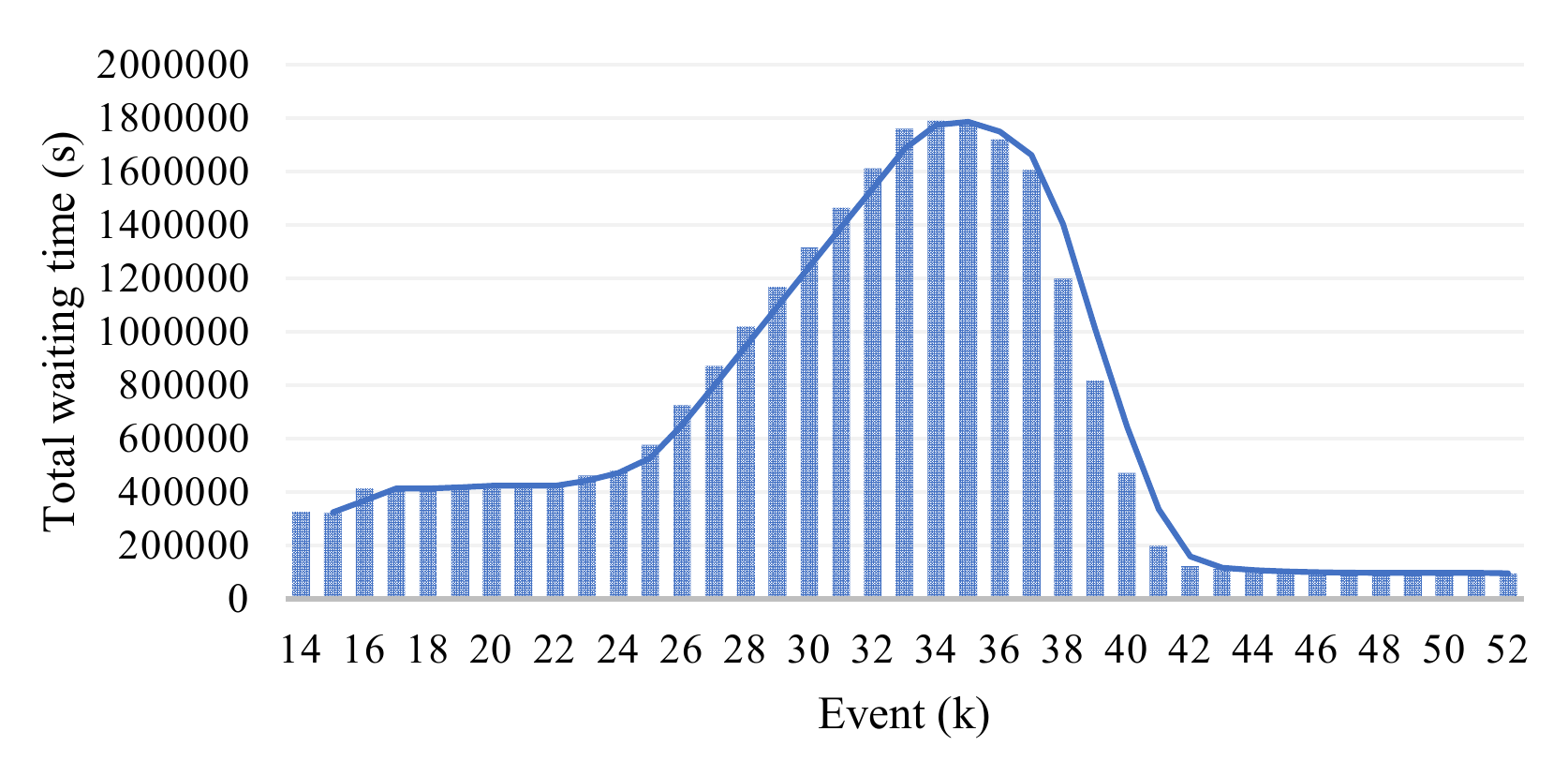}
  \caption{Passengers' total waiting time with  the STR method.}
  \label{fig.total waiting time}
  \vspace{-2mm}
\end{figure}

The results of  the number of on-board passengers and the number of stranded passengers at stations $8-12$ under  the proposed  STR method are plotted in  Figs.~\ref{fig.on-board passengers} and \ref{fig.stranded passengers}, respectively. From the two figures, we can observe that with the lasting large arriving rate and the limited train capacity, not all passengers can board the first coming train, and  some of them need to wait for next trains after event $k=24$. Meanwhile, as shown in  Fig.~\ref{fig.total waiting time}, the passenger total waiting time  increases dramatically. To avoid  a massive number of waiting  passengers accumulating at platforms, once the passengers exceed the allowable platform capacity, the  optimization procedure in  the PFM is triggered  and computes an optimal headway  under which the total waiting time of passengers is minimized, and the train load rate is maximized.\par
\begin{figure}[t]
  \centering\includegraphics[width=\hsize]{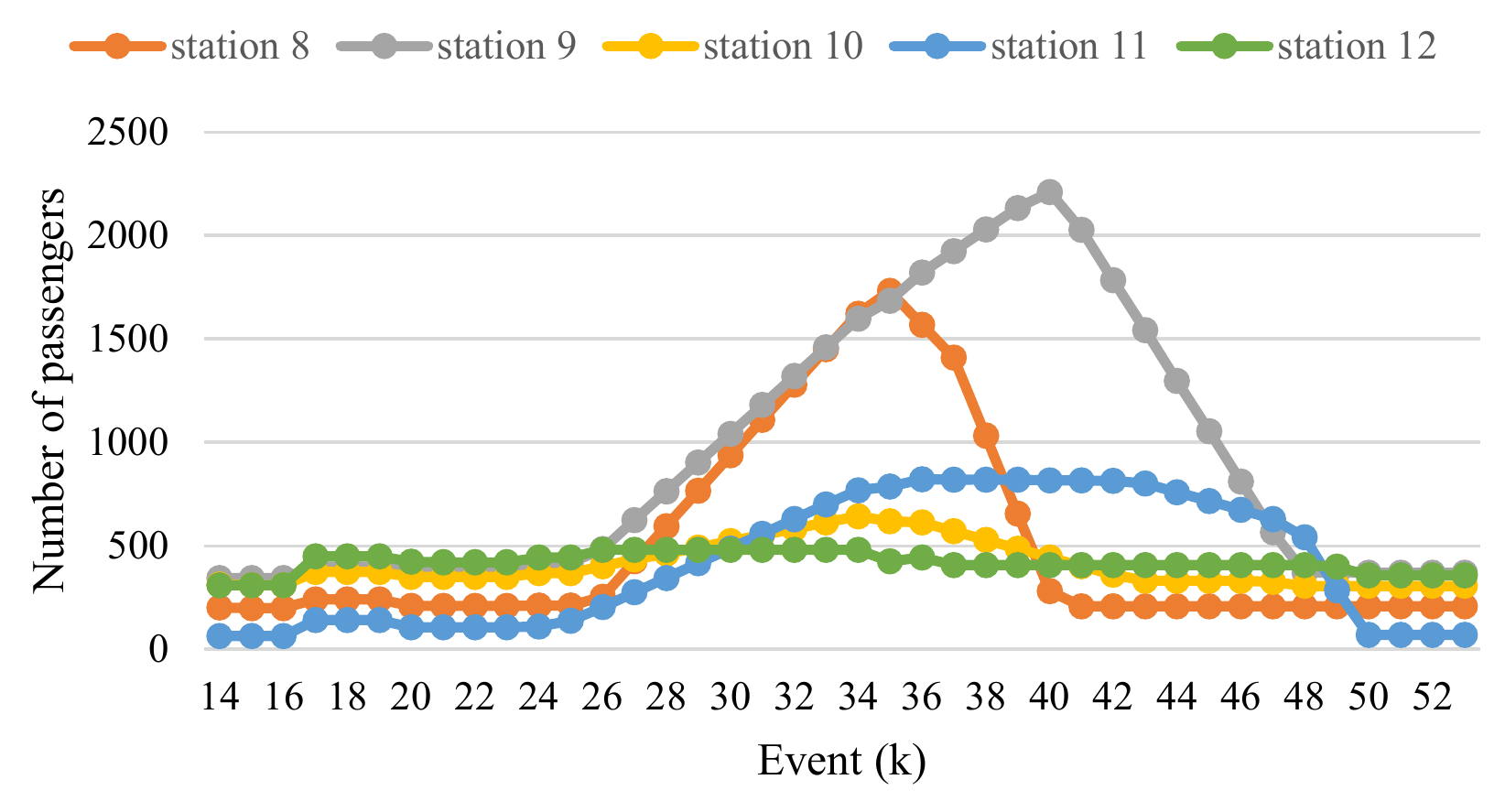}
  \caption{Maximal number of passengers on the platform with the fixed headway regulation method.}
  \label{fig.fixed maximal passengers}
\end{figure}
\begin{figure}[!htb]
  \centering\includegraphics[width=\hsize]{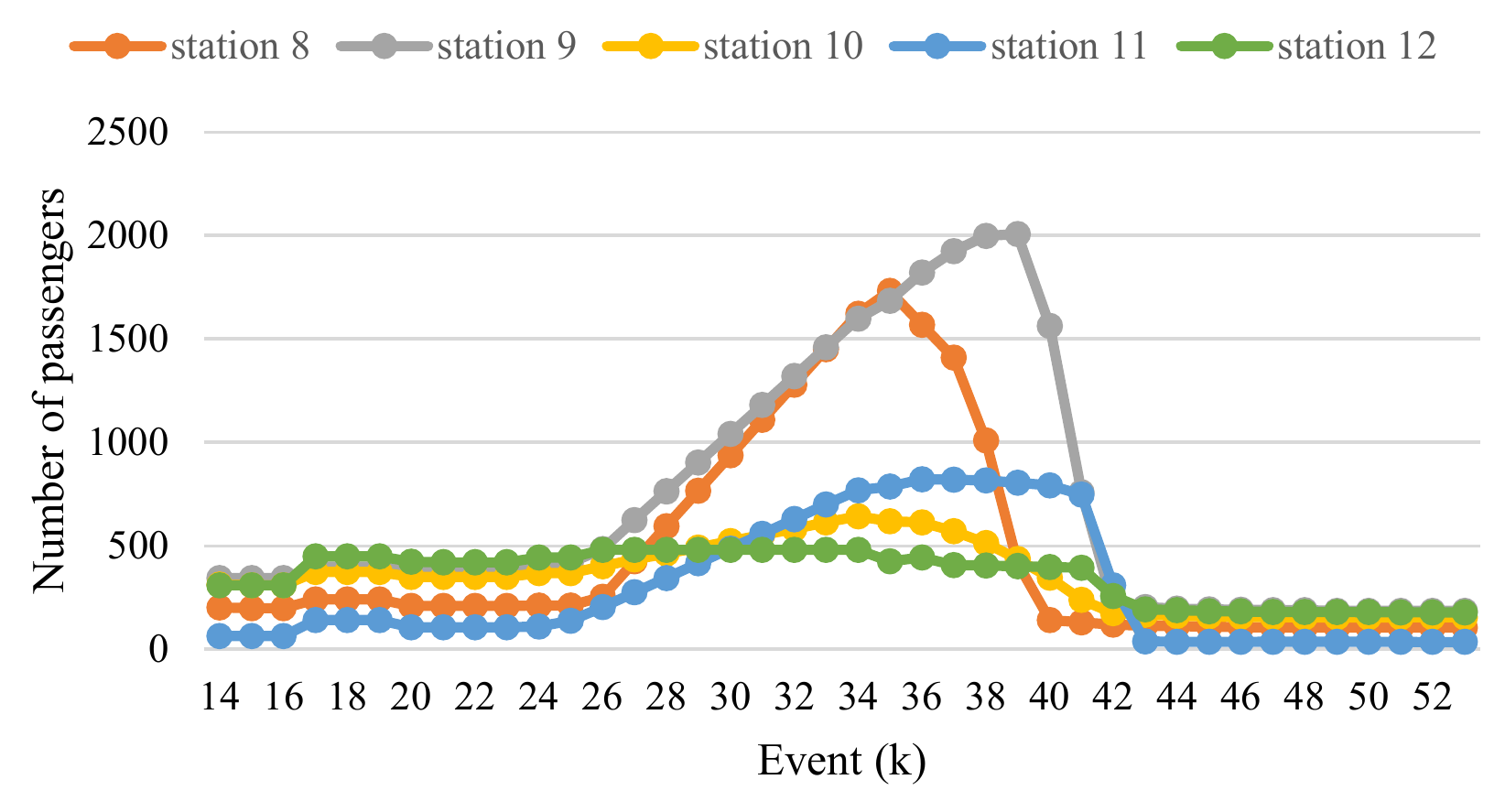}
  \caption{Maximal number of passengers on the platform with  the STR method.}
  \label{fig.maximal passengers}
\end{figure}
\begin{figure}[!htb]
  \centering\includegraphics[width=\hsize]{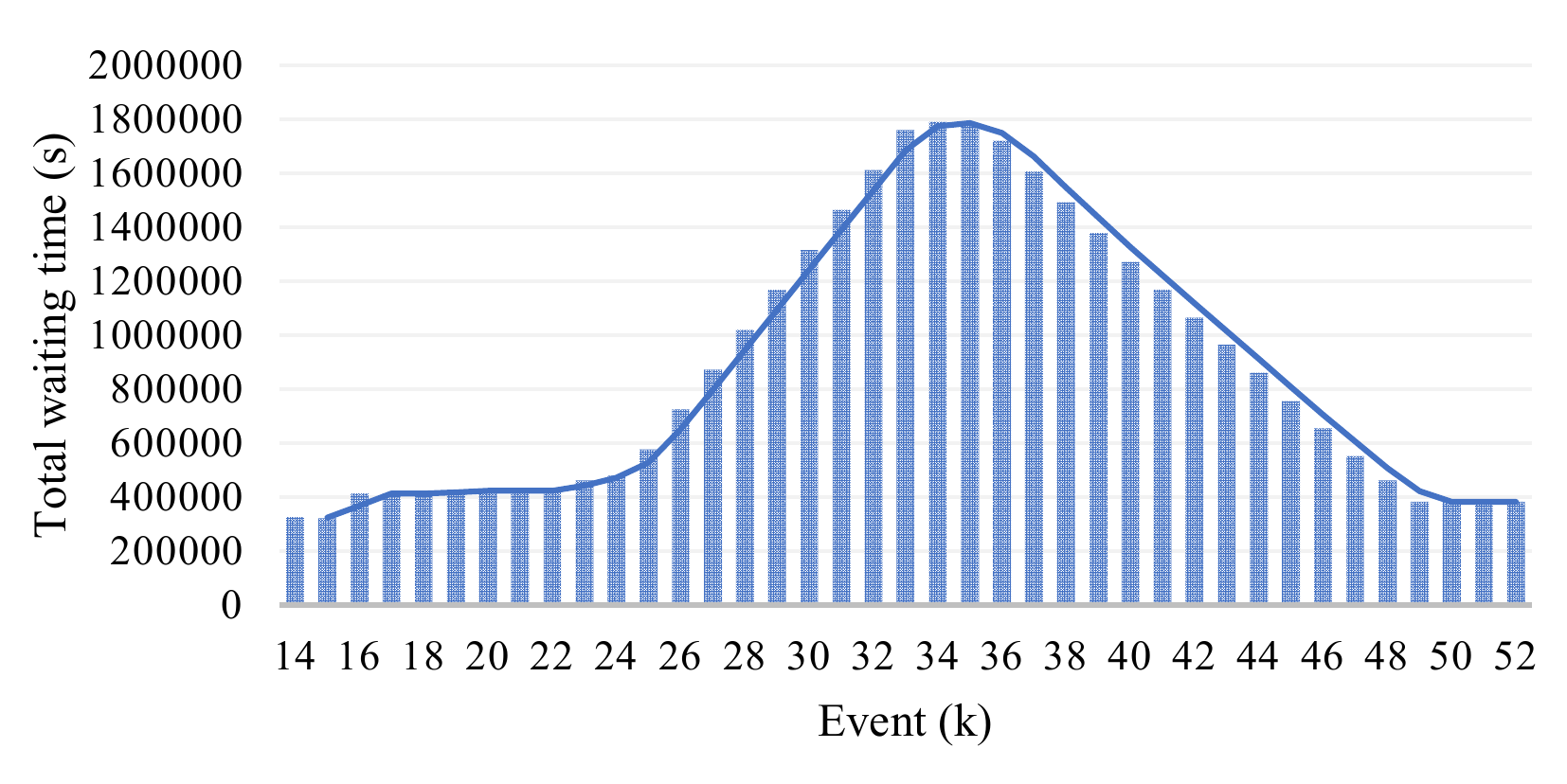}
  \caption{Passengers' total waiting time with the fixed headway regulation method.}
  \label{fig.fixed waiting time}
\end{figure}

 To better evaluate the performance, we compare the results with the case of using a fixed headway.  Figs.~\ref{fig.fixed maximal passengers} and~\ref{fig.fixed waiting time} respectively show the maximal number of passengers $p_{i,j}$ on the platform and the total waiting time of passengers in the case of not adjusting the headway, while Figs.~\ref{fig.maximal passengers} and \ref{fig.total waiting time} show the corresponding results of using the STR method. Comparing Figs.~\ref{fig.fixed maximal passengers} and \ref{fig.maximal passengers}, we can see that the maximal number of passengers is kept under 2000 by the STR method while it maximally reaches 2210  when the 31th train is at station 9 (i.e. at the 40th event). Moreover, the number of passengers at the platform declines faster after using the STR method and it reduces to a lower level after the 50th event.
From Figs.~\ref{fig.fixed waiting time} and \ref{fig.total waiting time}, it can be observed that the total waiting time  decreases gradually after the $35$th event and then keep  in a certain level after the period of large passenger flow.  However, using the STR method the total waiting time is much smaller, and at the meantime declines more quickly than that of using a fixed headway. 

\section{CONCLUSIONS}\label{conclusions}

In this work, a novel service-oriented metro traffic regulation method is  proposed to improve the operation performance under  the time-variant passenger flow and  the external disturbances. The presented method  integrates the train operation  module (TOM) together with  the passenger flow  module (PFM). When the number of passengers on platform is over a given threshold, the headway is recalculated to reduce the total waiting time of passengers. Meanwhile, the TOM takes the new headway to control trains to reduce the effect of disturbances. In this way, the traffic regulation problem can be effectively solved and the rescheduled train operation process will be more consistent with time-dependent passenger arrival demands. Numerical results of simulations conducted on the line 9 of the Beijing metro show that the total passenger waiting time, the number of stranded passengers are significantly reduced, meanwhile the load rate is kept in an optimal level compared to the use of fixed headway. In the future, the  enabling condition  for changing the headway back to the original one will be studied to improve automation and reduce the energy consumption of the service-oriented metro traffic regulation.



%
%
%

\end{document}